\documentclass[notitlepage]{article}

\usepackage{epsfig}

\begin{document}

\title{Geometrical picture of dynamical facilitation}

\author{Stephen Whitelam$^1$ and Juan P. Garrahan$^{1,2}$ \\ $^1${\em
Theoretical Physics, University of Oxford, 1 Keble Road, Oxford, OX1
3NP, UK} \\ $^2${\em School of Physics and Astronomy, University of
Nottingham, Nottingham, NG7 2RD, UK}}

\date{\today}

\maketitle

\begin{abstract}
Kinetically constrained models (KCMs) are models of glass formers
based on the concept of dynamic facilitation. This concept accounts
for many of the characteristics of the glass transition. KCMs usually
display a combination of simple thermodynamics and complex glassy
dynamics, the latter being a consequence of kinetic constraints.  Here
we show that KCMs can be regarded as systems whose configuration space
is endowed with a simple energy surface but a complicated geometry.
This geometry is determined solely by the kinetic constraints, and
determines the dynamics of the system. It does not affect the overall
distribution of states. Low temperature dynamical slow-down is then a
consequence of the competition between the paths allowed by the
geometry of configuration space, and those leading to energy
relaxation. This competition gives rise to dynamical bottlenecks
unrelated to static properties.  This view of the dynamics is distinct
from that based on an underlying static rugged energy landscape.  We
illustrate our ideas with simple examples.
\end{abstract}

\section{Introduction}

The view of the glass transition (see \cite{review1,review2,review3}
for general reviews) as a dynamical phenomenon emerges naturally from
the real-space picture of supercooled liquids developed in
\cite{Garrahan-Chandler,Berthier-Garrahan,Whitelam-Berthier-Garrahan,Berthier,Jung-Garrahan-Chandler}.
This work demonstrates that many of the properties of supercooled
liquids can be simply explained by the existence and statistical
properties of spatially correlated dynamics, or dynamical
heterogeneity \cite{DHreview1,DHreview2,DHreview3}.  Dynamical
heterogeneity has been observed both in deeply supercooled liquids
\cite{DHexp1}, in the mildly supercooled regime in simulations
\cite{DHnum}, and in experiments on colloidal suspensions
\cite{DHexp2}.

Heterogeneous dynamics, in combination with structural homogeneity, as
observed in glass formers, appears in systems possessing
both localized excitations and facilitated dynamics
\cite{Glarum,Palmer-et-al,Fredrickson-Andersen}.  The concept of
dynamic facilitation stipulates that the dynamics of a local region is
suppressed if is bordered by immobile regions, and thus reflects the
idea that particles in a jammed liquid cannot rearrange unless adjacent to mobile particles.  Dynamic facilitation is realized in
KCMs, which are simple models of non interacting
particles or excitations in a lattice whose dynamics is subject to
kinetic constraints (see \cite{Ritort-Sollich} for a review).
Examples of KCMs are spin facilitated models, like the
Fredrickson-Andersen (FA) model~\cite{Fredrickson-Andersen} or the
East model~\cite{Jackle}, and constrained lattice gases, such as the
Kob-Andersen model~\cite{Kob-Andersen}.

In this paper we study the effect that dynamic facilitation has on the
geometry of configuration space, and how in turn this affects the
dynamics \cite{nelson}.  We show, by generalizing KCMs to continuous
degrees of freedom, that kinetic constraints impose a metric structure
on configuration space which does not affect the static distribution
of states, and therefore the thermodynamics, but which has a dramatic
effect on the paths between states at low energies.  We show that
dynamical slow-down at low temperatures occurs because the system is
forced to evolve along configuration space paths of least metric
distance, or geodesics, which are distinct from the paths that relax
the energy.  This competition between available trajectories and
energy relaxation gives rise to dynamical barriers, or bottlenecks,
which are unrelated to the statics of the system.  This view of the
dynamics is distinct from that based on the idea that a static rugged
energy landscape is responsible for the properties of glass
formers (see e.g. \cite{review3}).

This paper is organised as follows. In section \ref{geometric} we
discuss the geometric interpretation of dynamic facilitation.  In
sections \ref{examples} and \ref{toy-model2} we illustrate our ideas
with simple examples.  In section \ref{conclusions} we state our
conclusions.

\section{Geometrical interpretation of constrained dynamics}
\label{geometric}

The real-space picture of supercooled liquids of Refs.\
\cite{Garrahan-Chandler,Berthier-Garrahan,Whitelam-Berthier-Garrahan,Jung-Garrahan-Chandler}
is based on two observations.  First, at low temperature very few
particles in the liquid are mobile, and these mobility excitations are
localized in space.  Second, regions of the liquid cannot become
mobile unless their neighbouring regions are mobile. These
observations are implemented as follows.  A supercooled fluid in $d$
spatial dimensions is coarse-grained into cells of linear size of the
order of the static correlation length as given by the pair
correlation function.  Cells are classified by a scalar mobility
field, $x$, identified by coarse-graining the system on a microscopic
time scale.  Mobile regions carry a free energy cost, and when
mobility is low interactions between cells are not important.
Adopting a thermal language, we expect static equilibrium to be
determined by the non-interacting Hamiltonian,
\begin{equation}
\label{hamiltonian}
H[x] = \sum_{\mu=1}^N f_\mu(x^\mu),
\end{equation}
where $\mu=1,\ldots,N$ labels the lattice site, $x^\mu$ is the
mobility field at site $\mu$, and $f_\mu$ is an analytic function of
$x^\mu$ chosen so that the ground state of the model is such that
$|x^\mu|$ is small.  At low mobility, the distinction between single
and multiple occupancy is irrelevant, as is the distinction between
discrete or continuous degrees of freedom.  In what follows we assume
$x^\mu$ to be real.  The dynamics of the mobility field is given by a
master equation,
\begin{equation}
\label{master1}
\partial_t P\left( x , t \right) =
\sum_\mu \mathcal{C}_\mu \left(  x  \right) \,
\hat{\mathcal{L}_\mu} \, P\left( x , t \right) ,
\end{equation}
where $P\left( x , t \right)$ is the probability that the system has
configuration $x \equiv \{x^\mu \,|\, \mu=1,2,...,N\}$ at time $t$.
The local operators $\hat{\mathcal{L}_\mu}$ are the same as those for
unconstrained local dynamics. Their action on $P$ can be written
\begin{equation}
\label{local}
\hat{\mathcal{L}_\mu} \, P\left( x \right) = \sum_{x'^{\mu}}
w(x'^{\mu} \rightarrow x^\mu) \, P\left(x'^{\mu},\{x^{\nu \neq
\mu}\}\right)- \sum_{x'^{\mu}} w(x^\mu \rightarrow
x'^{\mu}) \, P\left(x^\mu,\{x^{\nu \neq \mu}\}\right) ,
\end{equation}
where $w(x^\mu \rightarrow x'^{\mu})$ is the probability of going from
configuration $(x^\mu,\{x^{\nu \neq \mu}\})$ to configuration
$(x'^{\mu},\{x^{\nu \neq \mu}\})$ in unit time.  We ensure a unique
equilibrium configuration exists by requiring (\ref{master1}) to obey
detailed balance with respect to (\ref{hamiltonian}), i.e.
\begin{equation}
\frac{w(x\rightarrow x')}{w(x' \rightarrow x)}
=e^{-\beta(H[x']-H[x])}.
\label{db}
\end{equation} 
The kinetic constraint, $\mathcal{C}_\mu \left( x \right)$, is
designed to suppress the dynamics of cell $\mu$ when surrounded by
immobile regions.  It must also be such that Eq.\ (\ref{master1})
satisfies detailed balance, which is achieved if $\mathcal{C}_\mu
\left( x \right)$ does not depend on $x^\mu$ itself.  To reflect the
local nature of dynamic facilitation we allow $\mathcal{C}_\mu$ to
depend only on the $\{ x^\nu \}$ of nearest neighbours of $\mu$, and
require that $\mathcal{C}_\mu$ is small when local mobility is scarce.
Equations (\ref{hamiltonian})-(\ref{db}) define a generic KCM.

We now show that Eq.\ (\ref{master1}) has a simple geometrical
interpretation. Using standard methods \cite{Risken} we pass from the
master equation (\ref{master1}) to the Fokker-Planck equation
\begin{equation}
\label{fokkerplanck}
\partial_t P\left( x , t \right) =
\sum_\mu \mathcal{L}^{({\rm FP})}_\mu \,P \left( x , t \right) ,
\end{equation}
where
\begin{equation}
\label{fokkerplanck2}
\mathcal{L}^{({\rm FP})}_\mu = \frac{\partial}{\partial x^\mu} \left\{
\mathcal{C}_\mu \left( x \right) f_\mu'(x^\mu) + T
\frac{\partial}{\partial x^\mu} \mathcal{C}_\mu \left( x \right)
\right\}.
\end{equation}
The prime on $f_\mu$ denotes differentiation with respect to $x^\mu$.
Equation (\ref{fokkerplanck}) has a simple physical interpretation: it
describes single-particle driven diffusion on an $N$-dimensional
curved space $\{x^\mu \, |\, \mu = 1,2,...,N\}$ . The form of the
driving term in (\ref{fokkerplanck2}) comes from the Hamiltonian
(\ref{hamiltonian}) and the kinetic constraint.  But the curvature of
this space is due solely to the kinetic constraint: from the form of a
Fokker-Planck equation on a curved manifold~\cite{Risken} we identify
the inverse metric tensor of this space as $g^{\mu \nu}(x) =
\delta^{\mu \nu} \mathcal{C}_\mu \left( x \right)$.  Thus the metric
reads
\begin{equation}
\label{metric}
g_{\mu \nu}(x) = \delta_{\mu \nu} \mathcal{C}_\mu^{-1} \left( x  \right).
\end{equation} 
No summation is implied. The effect of dynamic facilitation is to
endow the configuration space $x$ with a nontrivial metric.  Since
dynamic facilitation is responsible for the interesting dynamical
properties of these models, we expect that these dynamical properties
can be extracted from the metric tensor.

The effect of the metric on configuration space trajectories is made
clearer by writing a formal solution to equation
(\ref{fokkerplanck})~\cite{ZinnJustin}:
\begin{equation}
\label{pathint}
P(x_b,t_b|x_a,t_a)=\int_a^b \sqrt{|g|} \, \mathcal{D}x \,
e^{-\mathcal{S}[x]},
\end{equation}
where $P(x_b,t_b|x_a,t_a)$ is the conditional probability that a
system starting at point $x_a$ in configuration space at time $t_a$ is
found at point $x_b$ at time $t_b$.  We have introduced the
determinant of the metric, $|g| \equiv \det \left( g_{\mu \nu} \right)
= \prod_{\mu} \mathcal{C}_{\mu}^{-1}$. The path integral
(\ref{pathint}) is taken over all configurations of the system,
weighted by the dynamic action $\mathcal{S}[x]$. Up to constant terms
this action reads
\begin{equation}
\label{action}
\mathcal{S}[x] = \frac{1}{T} \left( H[x_b]-H[x_a] \right) 
+ \frac{1}{T} \int_{t_a}^{t_b} dt\, \left(
\dot{x}^{\mu} g_{\mu \nu} \dot{x}^{\nu} + \frac{\partial
H}{\partial x^{\mu}} g^{\mu \nu} \frac{\partial H}{\partial x^{\nu}}
\right),
\end{equation}
where dots denote differentiation with respect to time, and the
summation convention for once-repeated upper and lower indices is
implied.

The action (\ref{action}) weights the configuration space trajectories
of our system.  The system's classical trajectory, i.e.\ the path
taken by the dynamics when $T \to 0$, is one of extremum
$\mathcal{S}$.  At finite $T$, stochastic fluctuations cause
perturbations about this path.  Equation (\ref{action}) shows clearly
the way the various ingredients of a dynamically facilitated model
influence its trajectory through configuration space.  The first term
depends only on the Hamiltonian (\ref{hamiltonian}) evaluated at the
endpoints.  It gives rise to Boltzmann weights in the path integral
(\ref{pathint}) and plays no role in choosing the trajectory of the
system.  The path is chosen by a competition between the two terms in
the integral of Eq.\ (\ref{action}).  The first term, $\mathcal{S}_1
\equiv T^{-1} \int dt \, \dot{x}^{\mu} g_{\mu \nu} \dot{x}^{\nu}$, is
purely geometrical.  It is the action for free motion in a curved
background.  It depends on the metric $g_{\mu \nu}$, and therefore on
$\mathcal{C}^{-1}$.  The second term, $\mathcal{S}_2 \equiv T^{-1}
\int dt \, \partial_{\mu} H \, g^{\mu \nu} \, \partial_{\nu} H$,
depends on the Hamiltonian (\ref{hamiltonian}), but is proportional to
the inverse metric $g^{\mu \nu}$, and so to $\mathcal{C}$.  At low
temperatures, we expect $\mathcal{S}_2$ to be important only far from
equilibrium, when mobility is plentiful---or equivalently, when the
magnitudes of the coordinates $|x^{\mu}|$ are large and $\mathcal{C}$
is large---and the system's relaxation will take place by way of
gradient descent on the energy surface.  When mobility has diminished
significantly the constraints $\mathcal{C}$ become small.  We then
expect $\mathcal{S}_1$ to dominate.  Hence the classical trajectory of
the system in the region of weak dynamic facilitation will be governed
by extremum $\mathcal{S}_1$.  This corresponds to the equation of a
geodesic~\cite{Misner-Thorne-Wheeler}:
\begin{equation}
\label{geodesic1}
\ddot{x}^{\mu}+\Gamma^{\mu}_{\alpha \beta} \dot{x}^{\alpha}
\dot{x}^{\beta}=0.
\end{equation}
The Christoffel symbols $\Gamma^{\mu}_{\alpha \beta}$ are defined as
$\Gamma_{\alpha \beta}^{\mu} \equiv \frac{1}{2} g^{\mu \nu} (g_{\nu
\alpha,\beta}+g_{\beta \nu,\alpha}-g_{\alpha \beta,\nu})$, and the
metric tensor raises and lowers indices.  Equation (\ref{geodesic1})
tells us how the kinetic constraint controls the system's trajectory
in configuration space in the dynamically constrained regime. When
mobility is scarce, the constraint $\mathcal{C}$ is small, and the
metric distance between points in configuration space, $s_{ab} \equiv
\int_a^b \sqrt{g_{\mu \nu} dx^{\mu} dx^{\nu}}$, is large, by virtue of
(\ref{metric}).  By minimising this distance, and so following
Eq. (\ref{geodesic1}), we expect the system to follow a path that does
not lead directly to a configuration of lowest energy.  This is a
geometrical interpretation of dynamical arrest.

\section{A simple example}
\label{examples}

We now illustrate these ideas with a simple example.  We define a
model of two variables, $\{ x^\mu \} = \{ x, y \}$, with each variable
constraining the other. Although such a model cannot exhibit true
glassiness it is useful because we can visualise directly the
two-dimensional configuration space it inhabits.  We choose the
Hamiltonian (\ref{hamiltonian}) to be quadratic,
\begin{equation}
H = \frac{J}{2} \left( x^2 + y^2 \right) ,
\end{equation}
where $J$ sets the energy scale.  We choose the constraint
functions to be quadratic also, with each coordinate constraining the other:
\begin{equation}
\mathcal{C}_x = y^2 , \;\;\; \mathcal{C}_y=x^2 .
\end{equation}
Thus, the Fokker-Planck equation (\ref{fokkerplanck}) is determined.
Equation (\ref{metric}) gives the metric of this system as
\begin{equation}
g_{\mu \nu}=\left(\begin{array}{cc} 1/y^2 & 0 \\ 0 & 1/x^2 \end{array}
\right).
\end{equation}
The metric is singular, and so distances in this configuration space
become very large, when $x$ or $y$ vanish, reflecting the tendency of
the kinetic constraint to suppress the dynamics of the system in the
regime of low mobility. The space described by this metric is
negatively curved with curvature scalar $R=-4 \left(x^2/y^2+y^2/x^2
\right)$ \cite{Misner-Thorne-Wheeler}.  The geodesics follow from
(\ref{geodesic1}) which in this case read
\begin{equation}
\label{geodesic2}
y\ddot{x}+(y/x)^3 \dot{y}^2-2 \dot{x} \dot{y}=0, \;\;\;
x \ddot{y}-(x/y)^3 \dot{x}^2-2 \dot{x} \dot{y}=0.
\end{equation}
To illustrate the effect of the kinetic constraint on the trajectories
chosen by the system we first cast Equation (\ref{fokkerplanck}) into
a form more convenient for numerical simulation. In the usual
way~\cite{VanKampen} we recognize that (\ref{fokkerplanck}) is
equivalent to the coupled Langevin equations
\begin{equation}
\label{langevin}
\left(\begin{array}{c}
\dot{x}(t)\\
\dot{y}(t)
\end{array} \right)=-
\left(\begin{array}{c}
J y^2(t) x(t)\\
J x^2(t) y(t)
\end{array} \right)+
\left(\begin{array}{c}
|y(t)|\eta_1(t)\\
|x(t)| \eta_2(t)
\end{array} \right),
\end{equation}
where the Gaussian white noise variables $\eta_i(t)$ have zero mean
and variance $\langle \eta_i(t) \eta_j(t') \rangle = 2T \delta(t-t')
\delta_{ij}$.  In this representation the kinetic constraint manifests
itself as a state dependent rate: the driving term is linear in
$\mathcal{C}$ while the noise term goes as $\sqrt{\mathcal{C}}$.  The
unconstrained version of the same model, i.e. one for which
$(\mathcal{C}_x, \mathcal{C}_y)=(1,1)$, would have Langevin equation
$\dot{x}^\mu(t)=-J x^\mu +\eta^\mu(t)$.

We now show explicitly that low temperature trajectories with fixed
initial and final configurations are approximated by the geodesic
equation (\ref{geodesic2}).  We have used Transition Path Sampling
(TPS) \cite{Tps} to ensure that we obtained a set of trajectories
which both obeyed the endpoint conditions are were well-sampled from
the dynamical action (\ref{action}).  In order to generate
out-of-equilibrium trajectories for the relaxation from high energy
configurations we used the Crooks-Chandler \cite{Crooks}
algorithm; alternatives such as the local algorithm of Ref.\
\cite{pratt} are also suitable.

\begin{figure}
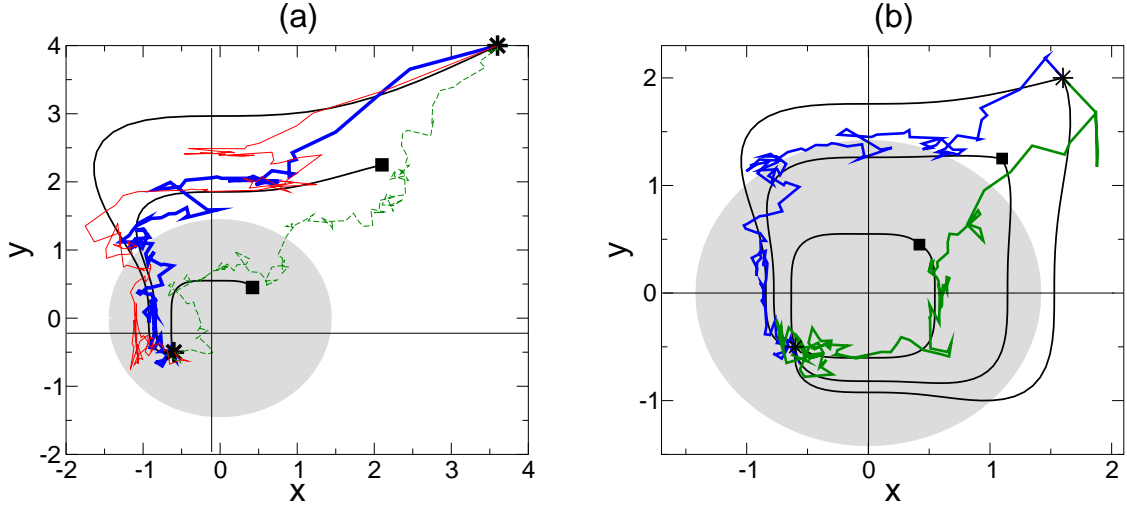

\begin{center}
\epsfig{file=fig1a.eps,width=7.8cm}
\epsfig{file=fig1b.eps,width=7.8cm}
\end{center}
\caption{ \label{fig1} Configuration space trajectories for the toy
model described in the text.  Panel (a):
$T=2.0,~J=1,~t_b-t_a=129,~(x_a, y_a)=(3.6,4.0),~(x_b,y_b)=(-0.6 \pm
0.01,-0.5 \pm 0.01)$.  We used TPS with the Crooks-Chandler
\cite{Crooks} algorithm to generate $2.78 \times 10^8$ trajectories,
starting from point $a$ (indicated by a star at the top right), of
which 11403 landed in the end-point $b$ (indicated by a star at the
bottom left).  The thin red path is a representative trajectory.  The
thick blue path is the mean of those trajectories.  The dashed green
path is the mean of 273 trajectories reaching $b$ from $a$ for an
unconstrained version of an otherwise identical model.  Numerical
solutions to the geodesic equation (\ref{geodesic2}) joining $b$ to
three other points (shown as black squares) are overlaid.  The gray
circle encloses the equilibrium region $\langle x^2 + y^2 \rangle =
T/J$.  Panel (b): a similar plot starting from $(x_a, y_a)=(1.6,2.0)$.
Again, $2.78 \times 10^8$ trajectories were generated.  The thick blue
path shows the mean of 2199 trajectories landing in region $b$ which
follow the left hand side route.  The average path over 6 right hand
side trajectories is also shown in green.}
\end{figure}

Figure \ref{fig1} presents the comparison of the dynamical evolution
of our toy KCM with the geometry of its configuration space.  TPS was
used on the system defined by Eq.\ (\ref{langevin}) to generate
trajectories between the configurations indicated by a star.  In Fig.\
1a we show both a characteristic trajectory (thin red path) and the
average over the set of all trajectories obtained via TPS (thick blue
path).  We also show geodesics (black curves) joining the final
configuration to the initial configuration (indicated by a star), and
to other configurations along the gradient descent path (indicated by
black squares).  The solutions to (\ref{geodesic2}) are degenerate:
there is one geodesic going left from the initial point and another
going right. The distinction depends on the initial conditions on the
path.  In the TPS simulations we force the initial displacement to be
to the left to obtain a set of trajectories which follows the left
hand geodesics.  These are the paths shown in Fig.\ 1a and 1b.  In
Fig.\ 1b we also show the average over a small set of right hand TPS
trajectories and the corresponding pairs of geodesics.

In Fig.\ \ref{fig1} we see the features described in the previous
section.  Initially the system is in the high energy region and it
relaxes approximately by gradient descent.  As it reaches the region
of weak dynamic facilitation at low energies, the system relaxes by
following approximately the geodesics of configuration space.  Shown
for comparison in Fig.\ 1a are the trajectories followed by an
unconstrained version of the same model (dashed green path).  In this
case relaxation is by gradient descent.  Given that the rate for
motion in the $x$ (resp.\ $y$) direction vanishes when $y$ (resp.\
$x$) vanishes, trajectories have to cross the coordinate axis
orthogonally, which gives trajectories (and geodesics) a
``rectangular'' shape.  Equipotential lines are concentric circles,
and the conflict between allowed dynamical paths and energy relaxation
is evident.

\section{An even simpler example}
\label{toy-model2}
We can gain more insight into the nature of dynamic facilitation by
considering an even simpler, one-variable model.  This can be thought
of as modeling facilitated dynamics for an order parameter, rather
than for microscopic variables.  Equations (\ref{fokkerplanck}) and
(\ref{fokkerplanck2}) tell us that an $N$-variable, dynamically
constrained system looks like a single particle diffusing in $N$ space
dimensions.  Thus a one-variable dynamically constrained model looks
like a single particle diffusing in one space dimension. We can
exploit the fact that one dimensional spaces are always
flat~\cite{Misner-Thorne-Wheeler} to `remove' the kinetic constraint.
We will see below that this can be done at the expense of introducing
an effective free energy barrier.

Consider a model with one degree of freedom, $x$, with Hamiltonian
(\ref{hamiltonian}) $H=J x^2/2$. We choose the dynamical constraint to
be $\mathcal{C}(x)$.  This differs from the higher-dimensional cases in
that the variable constrains itself.  Consequently, the transformation
from Fokker-Planck to Langevin form is ambiguous~\cite{VanKampen}, and
depends on whether one choses pre-point or mid-point discretisation of
time-dependent quantities (It\^{o} or Stratonovich calculus). We choose
the former, so that our model is causal.  The Langevin description of
our one-variable model is then
\begin{equation}
\label{onevar}
\dot{x}(t)=-J \, x \, \mathcal{C}(x)+T \mathcal{C}'(x)+\sqrt{\mathcal{C}(x)} \eta(t),
\end{equation}
where the prime denotes differentiation with respect to $x$, and the
noise term $\eta(t)$ has zero mean and variance $\langle \eta(t)
\eta(t') \rangle = 2T \delta(t-t')$.  Since in this case the
constraint depends on the variable it constrains, the second term on
the right hand side of (\ref{onevar}) is needed to ensure an equilibrium
density proportional to $\exp{( - J x^2 / 2 T)}$ \cite{FP}.  In the
higher-dimensional cases we considered above there was no
self-facilitation and this term was absent.  The corresponding
unconstrained model has $\mathcal{C}(x)=1$, and its Langevin equation
is $\dot{x}(t)=-\partial_x (J x^2 / 2 )+\eta(t)$.

The appearance of the multiplicative noise term $\sqrt{\mathcal{C}(x)}
\eta(t)$ in Eq.\ (\ref{onevar}) is due to the metric-augmented
diffusion term in (\ref{fokkerplanck2}), and is therefore a
consequence of the kinetic constraint. The statement that a
one-dimensional space is flat is equivalent to the statement that the
noise may be made additive by a local change of variables. If we make
the change of variables
\begin{equation}
\label{change-of-variables}
y(t) \equiv \int^{x(t)} \frac{dx'}{\sqrt{\mathcal{C}(x')}},
\end{equation} 
we can use It\^{o}'s formula~\cite{VanKampen} to rewrite
(\ref{onevar}) as
\begin{equation}
\label{onevar2}
\dot{y}(t)=-J x \sqrt{\mathcal{C}(x)}+ \frac{1}{2} \, T \,
\sqrt{\mathcal{C}(x)} \, \partial_x \ln \mathcal{C}(x) +\tilde{\eta}(t),
\end{equation}
where $x$ should be eliminated in favour of $y$. The noise
$\tilde{\eta}(t)$ has the same mean and variance as $\eta(t)$.
Provided we can write the deterministic terms in (\ref{onevar2}) as
the gradient of an effective potential we can regard Equation
(\ref{onevar2}) as unconstrained, in the sense that the prefactor of
the noise does not contain the dependent variable.

\begin{figure}
\begin{center}
\epsfig{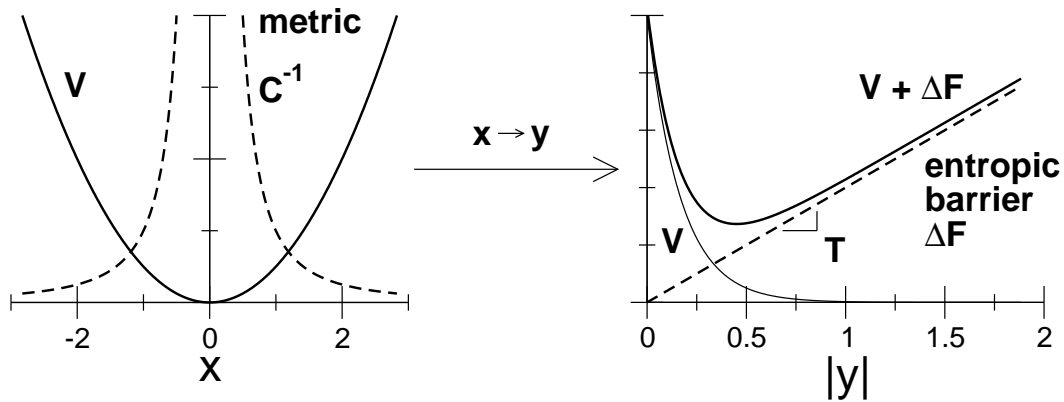}
\end{center}
\caption{ \label{fig2} An illustration of how a constrained problem may
have an alternative representation as an unconstrained one with an
extra free energy barrier (from the example of section
\ref{toy-model2}).  The constraint ${\cal C}(x)$ fixes the metric
$g(x)={\cal C}^{-1}(x)$ of space $x$.  This implies that metric
distances become large in the region of small $x$.  Under the
transformation Eq.\ (\ref{change-of-variables}) the problem changes
into a non constrained one, $g(y)=1$, but with an extra free energy
barrier $\Delta F(y)$ which excludes the system from the low energy
region of large $|y|$. }
\end{figure}

Consider the specific case of a quadratic constraint,
$\mathcal{C}(x)=x^2$.   Equation (\ref{onevar}) becomes
\begin{equation}
\dot{x} = - J \, x^3 + 2 \, T \, x + x \, \eta,
\end{equation}
where we have assumed $x>0$.  The change of variables
(\ref{change-of-variables}) now amounts to a Cole-Hopf transformation,
$y \equiv \ln x$.  Hence the regime where dynamic facilitation is
relevant corresponds to $y \ll 0$.  Equation (\ref{onevar2}) now
reads
\begin{equation}
\label{unconstrained}
\dot{y}=- \partial_y \left( \frac{J}{2} \, e^{2 y} - T \, y \right)
+ \tilde{\eta}.
\end{equation}
Equation (\ref{unconstrained}) looks like an unconstrained model with
an effective {\em entropic} barrier $\Delta F = - T y = - T \, \ln x$,
which prevents the particle from accessing the dynamically constrained
regime of $y \ll 0 \Rightarrow x \sim 0$ \cite{FPy}.  Figure
\ref{fig2} illustrates this.

\section{Conclusions}  \label{conclusions}

In this paper we have presented a geometric interpretation of dynamic
facilitation.  By generalizing kinetically constrained models to
continuous degrees of freedom we have shown that in these systems the
kinetic constraints can be seen as endowing configuration space with a
non-trivial metric structure.  This metric structure in turn
determines the dynamical trajectories.  In the regime where the
kinetic constraint is dominant, i.e.\ at low temperatures or energies,
the geometry forces the dynamics to proceed through paths close to
geodesics, rather than close to paths which relax the energy.  This
conflict between available paths and energy relaxation gives rise to
dynamical bottlenecks or barriers, despite the fact that there are no
static barriers in these systems.

Our view of dynamical arrest in terms of the geometry of configuration
space is distinct from that based on the idea of motion in an
effective rugged energy (or free energy) landscape (see \cite{review3}
and references therein).  In the landscape perspective it is often
assumed that relevant features of the energy surface, like local
minima and saddle points, determine the behaviour of the system, and
that knowledge of the statistical properties of these features is
sufficient to explain both dynamic and thermodynamic properties.  In
our approach the effective energy surface displays no particular
features, the thermodynamics is uninteresting, and there are no
special configurations, such as local minima or saddles, which play
any major roles either in the statics or in the dynamics.  All the
interesting structure is in the paths between configurations, and
therefore in the dynamics. In this picture it is the effective metric
structure of configuration space that distinguishes a glassy system
from a normal one, and not the ruggedness of its energy surface.

\bigskip 
We are grateful to D. Chandler for discussions.  We acknowledge
financial support from EPSRC Grants No.\ GR/R83712/01 and
GR/S54074/01, the Glasstone Fund, and Linacre College Oxford.


\begin{thebibliography}{99}
\bibitem{review1} C.A. Angell, Science {\bf 267}, 1924 (1995).

\bibitem{review2} M.D. Ediger, C.A. Angell and S.R. Nagel,
J. Phys. Chem. {\bf 100}, 13200 (1996).

\bibitem{review3} P.G. Debenedetti and F.H. Stillinger, Nature {\bf
410}, 259 (2001).

\bibitem{Garrahan-Chandler} J.P. Garrahan and D. Chandler,
Phys. Rev. Lett. {\bf 89}, 035704 (2002); Proc. Natl. Acad. Sci. USA
{\bf 100}, 9710 (2003).
 
\bibitem{Berthier-Garrahan} L. Berthier and J.P. Garrahan,
J. Chem. Phys. {\bf 119}, 4367 (2003); Phys. Rev. E {\bf 68}, 041201
(2003).

\bibitem{Whitelam-Berthier-Garrahan} S. Whitelam, L. Berthier, and
J.P. Garrahan, e-print cond-mat/0310207 (2003).

\bibitem{Berthier} L. Berthier, e-print cond-mat/0310210 (2003).

\bibitem{Jung-Garrahan-Chandler} Y. Jung, J.P. Garrahan and
D. Chandler, e-print cond-mat/0311396 (2003).

\bibitem{DHreview1} H. Sillescu, J. Non-Cryst. Solids {\bf 243}, 81
(1999).

\bibitem{DHreview2} M.D. Ediger, Annu. Rev. Phys. Chem.  {\bf 51}, 99
(2000).

\bibitem{DHreview3} S.C. Glotzer, J. Non-Cryst. Solids, {\bf 274},
342 (2000).

\bibitem{DHexp1} K. Schmidt--Rohr and H. Speiss, Phys. Rev. Lett.
{\bf 66}, 3020 (1991); M.T. Cicerone and M.D. Ediger,
J. Chem. Phys. {\bf 103}, 5684 (1995); E.V. Russell et al.,
Phys. Rev. Lett. {\bf 81}, 1461 (1998); L.A. Deschenes and D.A. Vanden
Bout, Science {\bf 292}, 255 (2001).

\bibitem{DHnum} T. Muranaka and Y. Hitawari, Phys. Rev. E {\bf 51},
R2735 (1995); D. Perera and P. Harrowell, Phys. Rev. E {\bf 51}, 314
(1995); B. Doliwa and A. Heuer, Phys. Rev. Lett. {\bf 80}, 4915
(1998); C. Donati et al., Phys. Rev. E {\bf 60}, 3107 (1999).

\bibitem{DHexp2} E. Weeks et al., Science {\bf 287}, 627 (2000).

\bibitem{Glarum} S.H. Glarum, J. Chem. Phys. {\bf 33}, 639 (1960).

\bibitem{Palmer-et-al} R.G Palmer, D.L. Stein, E. Abrahams and
P.W. Anderson, Phys. Rev. Lett. {\bf 53}, 958 (1984).

\bibitem{Fredrickson-Andersen} G.H. Fredrickson and H.C. Andersen,
Phys. Rev. Lett. {\bf 53}, 1244 (1984).

\bibitem{Ritort-Sollich} F. Ritort and P. Sollich, Adv. in Phys. {\bf
52}, 219 (2003).

\bibitem{Jackle} J. J\"{a}ckle and S. Eisinger, Z. Phys. {\bf B84},
115 (1991).

\bibitem{Kob-Andersen} W. Kob and H.C. Andersen, Phys. Rev. E {\bf
48}, 4364 (1993).

\bibitem{nelson} For other perspectives on glassiness also based on
curved spaces, but with a different microscopic origin, see for
example: P.J. Steinhardt, D.R. Nelson and M. Ronchetti,
Phys. Rev. Lett. {\bf 47}, 1297 (1981) and D.R. Nelson, Phys. Rev. B
{\bf 28}, 5515 (1983).

\bibitem{Risken} H. Risken, ``The Fokker-Planck equation: Methods of
solution and Applications'', (Springer Verlag, New York, 1984).

\bibitem{ZinnJustin} J. Zinn-Justin, ``Quantum Field Theory and
Critical Phenomena'', (OUP, Oxford 1989).

\bibitem{Misner-Thorne-Wheeler} C.W. Misner, K.S. Thorne, and
J.A. Wheeler, Gravitation, W.H. Freeman and Company, San Francisco
(1973).

\bibitem{VanKampen} N.G. Van Kampen, ``Stochastic Processes in Physics
and Chemistry'', (North-Holland, Amsterdam, 2001).

\bibitem{Tps} P.G. Bolhuis, D. Chandler, C. Dellago and P.L. Geissler,
Ann. Rev. Phys. Chem. {\bf 59}, 291 (2002).

\bibitem{Crooks} G.E. Crooks and D. Chandler, Phys. Rev. E {\bf 64},
026109 (2001).

\bibitem{pratt} L.R. Pratt, J. Chem. Phys. {\bf 85}, 5045 (1986).

\bibitem{FP} In the It\^o convention, the Fokker-Planck equation for
the Langevin equation (\ref{onevar}) is $\partial_t P\left( x , t
\right) = \partial_x \left[ J \, x \, \mathcal{C}(x) \, P\left( x , t
\right) - T \mathcal{C}'(x) \, P\left( x , t \right) \right] + T \,
\partial^2_x \left[ \mathcal{C}(x) \, P\left( x , t \right) \right]$.
The right hand side of this equation can be rewritten as $\partial_x
\left\{ \mathcal{C}(x) \left[ J \, x \, P\left( x , t \right) + T
\partial_x P\left( x , t \right) \right] \right\}$, leading to the
equilibrium density $P_{\rm eq.}(x) \propto \exp{( - J x^2 / 2 T)}$,
which is indeed independent of the constraining function
$\mathcal{C}(x)$.  If the Stratonovich convention is used, the second
term in the right hand side of Eq.\ (\ref{onevar}) is then preceded by
a factor $1/2$, giving the same Fokker-Planck equation as above.

\bibitem{FPy} The equilibrium density in the $y$ representation is
$P_{\rm eq.}(y) \propto \exp{( - J e^{2y} / 2 T + y)}$.  This
corresponds to the density $P_{\rm eq.}(x)$ after the change of
variables $x \to y=\ln{x}$, as can be seen from the invariance of the
partition function: $Z = \int \exp{( - J x^2 / 2 T)} \, dx = \int
\exp{( - J e^{2y} / 2 T)} \, e^{y} \, dy$.

\end{thebibliography}
\end{document}